\documentclass{emulateapj}
\usepackage{times,mathptm}
\usepackage{lscape}

\usepackage{epsf}
\usepackage{psfig}
\usepackage{lscape}

\def\etal{{et\,al.\,}}

\def\ltsima{$\; \buildrel < \over \sim \;$}
\def\simlt{\lower.5ex\hbox{\ltsima}}
\def\gtsima{$\; \buildrel > \over \sim \;$}
\def\simgt{\lower.5ex\hbox{\gtsima}}

\journalinfo{The Astrophysical Journal, 2008, arXiv:0803.0636}

\begin{document}
\title{Reliable Identification of Compton-thick Quasars at $z\approx$~2:\\
{\it Spitzer} Mid-Infrared Spectroscopy of HDF-oMD49}

\author{D.M. Alexander,\altaffilmark{1} R.-R. Chary,\altaffilmark{2}
  A. Pope,\altaffilmark{3,4,5} F.E. Bauer,\altaffilmark{6}
  W.N. Brandt,\altaffilmark{7} E. Daddi,\altaffilmark{8}
  M. Dickinson,\altaffilmark{5}\\
  D. Elbaz,\altaffilmark{8} and
  N.A. Reddy\altaffilmark{5}}

\affil{$^1$Department of Physics, Durham University, Durham DH1 3LE, UK}
\affil{$^2${\it Spitzer} Science Center, Caltech, MS 220-6, CA 91125, USA}
\affil{$^3${\it Spitzer} Fellow}
\affil{$^4$Department of Physics \& Astronomy, University of British Columbia, Vancouver, BC, V6T 1Z1, Canada}
\affil{$^5$National Optical Astronomy Observatory,  950 N. Cherry Ave., Tucson, AZ, 85719, USA}
\affil{$^6${\it Chandra} Fellow; Columbia Astrophysics Laboratory, Columbia University, Pupin
  Laboratories, 550 West 120th Street, New York, NY 10027, USA}
\affil{$^7$Department of Astronomy and Astrophysics, The Pennsylvania State University, 525 Davey Lab, University Park, PA 16802, USA}
\affil{$^8$Laboratoire AIM, CEA/DSM - CNRS - Universit\`e Paris Diderot, Irfu, Orme des Merisiers, 91191 Gif-sur-Yvette, France}

\shorttitle{RELIABLE IDENTIFICATION OF DISTANT COMPTON-THICK QUASARS}

\shortauthors{ALEXANDER ET AL.}

\slugcomment{Received 2005 Mar 10; accepted 2008 July 10}

%
\begin{abstract} 
%

Many models that seek to explain the origin of the unresolved X-ray
background predict that Compton-thick Active Galactic Nuclei (AGNs) are
ubiquitious at high redshift. However, few distant Compton-thick AGNs
have been reliably identified to date.  Here we present {\it
Spitzer}-IRS spectroscopy and 3.6--70~$\mu$m photometry of a $z=2.211$
optically identified AGN (HDF-oMD49) that is formally undetected in the
2~Ms {\it Chandra} Deep Field-North (CDF-N) survey. The {\it
Spitzer}-IRS spectrum and spectral energy distribution of this object
is AGN dominated, and a comparison of the energetics at X-ray
wavelengths to those derived from mid-infrared (mid-IR) and optical
spectroscopy shows that the AGN is intrinsically luminous ($L_{\rm
2-10~keV}\approx3\times10^{44}$~erg~s$^{-1}$) but heavily absorbed by
Compton-thick material ($N_{\rm H}\gg10^{24}$~cm$^{-2}$); i.e.,\ this
object is a Compton-thick quasar. Adopting the same approach that we
applied to HDF-oMD49, we found a further six objects at
$z\approx$~2--2.5 in the literature that are also X-ray weak/undetected
but have evidence for AGN activity from optical and/or mid-IR
spectroscopy, and show that all of these sources are likely to be
Compton-thick quasars with $L_{\rm 2-10 keV}>10^{44}$~erg~s$^{-1}$. On
the basis of the definition of Daddi \etal (2007), these Compton-thick
quasars would be classified as mid-IR excess galaxies, and our study
provides the first spectroscopic confirmation of Compton-thick AGN
activity in a subsample of these $z\approx$~2 mid-IR bright
galaxies. Using the four objects that lie in the CDF-N field, we
estimate the space density of reliably identified Compton-thick quasars
[\hbox{$\Phi\approx$~(0.7--2.5)}~$\times10^{-5}$~Mpc$^{-3}$ for $L_{\rm 2-10
keV}>10^{44}$~erg~s$^{-1}$ objects at $z\approx$~2--2.5] and show that
Compton-thick accretion was probaby as ubiquitious as unobscured
accretion in the distant Universe.
\end{abstract}




%
\section{Introduction}
%

There is a growing need for a complete census of Active Galactic Nuclei
(AGNs). The seminal discovery that every massive galaxy in the local
Universe harbors a super-massive black hole ($M_{\rm
BH}\simgt10^{6}$~$M_{\odot}$) implies that all massive galaxies must
have hosted AGN activity at some time during the past $\approx$~13~Gyrs
(e.g.,\ Rees 1984; Kormendy \& Richstone 1995). To trace accurately how
and when these black holes grew requires a detailed census of AGN
activity that will provide, amongst other things, estimates of the
efficiency and duty cycle of black-hole growth and any dependencies of
nuclear obscuration on AGN luminosity, redshift, and host-galaxy type
(e.g.,\ Marconi \etal 2004; La Franca \etal 2005; Shankar \etal 2007).

The exceptional sensitivity of the {\it Chandra} Deep Field
observations (e.g.,\ Brandt et~al. 2001; Giacconi et~al. 2002;
Alexander et~al. 2003a) has helped to unveil a $>10$ times larger
population of AGNs than found at most other wavelengths
($\approx$~7200~deg$^{-2}$; e.g.,\ Bauer et~al. 2004). Optical
spectroscopic follow-up observations have shown that these AGNs are
detected out to $z\approx$~5, and detailed X-ray spectral analyses have
revealed that the majority of the sources are obscured by gas and dust
(e.g.,\ Barger et~al. 2003; Szokoly et~al. 2004; Tozzi et~al. 2006; see
Brandt \& Hasinger 2005 for a review). However, although clearly
effective at identifying even heavily obscured AGNs out to high
redshift, there is compelling evidence that a large fraction of the AGN
population remains undetected at the $<10$~keV observed-frame energies
probed by these surveys: (1) about half of the X-ray background (XRB)
is unresolved at $>6$~keV (Worsley et~al. 2005), (2) the {\it observed}
obscured:unobscured AGN ratio is lower than that found for comparably
luminous AGNs in the local Universe (e.g.,\ Treister \& Urry 2005), (3)
the \hbox{5--10~keV} blank-field number counts are steeply rising at
the faintest X-ray fluxes (Rosati et~al. 2002), and (4) few
Compton-thick AGNs ($N_{\rm H}\simgt1.5\times10^{24}$~cm$^{-2}$) have
been identified, even though they comprise $\approx$~50\% of the AGN
population in the local Universe (e.g.,\ Risaliti et~al. 1999;
Guainazzi et~al. 2005; Tozzi et~al. 2006).

Many of the X-ray undetected AGNs are expected to be intrinsically
luminous sources that are heavily obscured by Compton-thick material
(i.e.,\ $N_{\rm H}>1.5\times10^{24}$~cm$^{-2}$; see Comastri 2004 for a
review). The most robust identification of a Compton-thick AGN is made
from high signal-to-noise ratio (S/N) X-ray spectroscopy, where the
detection of a high equivalent-width Fe~K emission line
($W_{\lambda}>$~1~keV at rest-frame energies 6.4--6.9~keV) and a
steeply rising reflection component at $>10$~keV reveals that little or
no X-ray emission is being seen directly, implying that the central
source is very heavily absorbed (e.g.,\ George \& Fabian 1991; Matt,
Brandt, \& Fabian 1996; Maiolino et~al. 1998; Matt
et~al. 2000). However, since the extreme obscuration toward the nucleus
of a Compton-thick AGN renders the observed $<10$~keV emission
$\approx$~30--1000 times fainter than the intrinsic (i.e.,\ unabsorbed)
emission, it is often challenging to identify robustly these sources
on the basis of their X-ray data alone; the range of absorption
correction factors is based on the observed to intrinsic X-ray
luminosity ratio for the AGNs in Table~8.1 of Comastri (2004), with the
observed X-ray fluxes from Bassani et~al. (1999). For example, despite
Compton-thick AGNs comprising half of the AGN population in the local
Universe, only $\approx$~50 sources have been reliably identified from
X-ray data to date (e.g.,\ Comastri 2004). Fortunately, although not as
conclusive as high S/N X-ray spectroscopy, the presence of
Compton-thick absorption can also be identified in X-ray weak AGNs when
a reliable measurement of the intrinsic power of the AGN is available
at other wavelengths; we stress here the necessity for X-ray
constraints in order to identify a Compton-thick AGN since it is only
in the X-ray band that an absorbing column density can be measured or
inferred.

Two of the most reliable measurements of the intrinsic power of an
obscured AGN are the luminosity of the mid-infrared (mid-IR; rest-frame
$>3$~$\mu$m; e.g.,\ Efstathiou \& Rowan-Robinson 1995; Granato
et~al. 1997; Lutz et~al. 2004) continuum and high-excitation emission
lines (i.e.,\ [OIII]$\lambda$5007; e.g.,\ Allen 1973; Kwan \& Krolik
1981; Bassani et~al. 1999), both of which are believed to be directly
illuminated by the central source. These emission regions provide a
good proxy for the intrinsic AGN luminosity, even in the presence of
extreme Compton-thick absorption, since they are more extended than the
X-ray absorbing material (i.e.,\ larger than the broad-line region;
e.g.,\ Lamer et~al. 2003; Risaliti et~al. 2007). As the luminosities of
both the mid-IR continuum and the high-excitation emission-line region
are dependent on many factors, including the location and geometry of
the region with respect to the central source, more robust constraints
are placed if both measurements are available.

Comprehensive evidence for a large X-ray undetected AGN population has
been found from a variety of X-ray based analyses using sources
detected in deep {\it Spitzer} surveys at mid-IR wavelengths,
suggesting that Compton-thick AGNs may be ubiquitious in the distant
Universe (e.g.,\ Donley et~al. 2005, 2007; Alonso-Herrero et~al. 2006;
Polletta et~al. 2006; Daddi et~al. 2007a; Steffen et~al. 2007;
Mart{\'{\i}}nez-Sansigre et~al. 2007; Fiore et~al. 2008). However, all
of these studies have relied on photometric observations for the
identification of candidate Compton-thick AGNs, leading to significant
uncertainties in measurements of the intrinsic AGN luminosity, the
absorbing column density, the degree of contamination from
star-formation activity, and by implication, the space density of
distant Compton-thick AGNs. In this paper we present {\it Spitzer}-IRS
spectroscopy and 3.6--70~$\mu$m observations of an optically identified
narrow emission-line AGN at $z=2.211$ (HDF-oMD49; Steidel et~al. 2002)
that is undetected in the published catalogs of the deepest X-ray
observation currently available [the 2~Ms {\it Chandra} Deep
Field-North (CDF-N); Alexander et~al. 2003a]. From a comparison of the
AGN energetics at X-ray wavelengths to those determined from optical
and mid-IR spectroscopy, we robustly show that HDF-oMD49 hosts an
intrinsically luminous ($L_{\rm 2-10
keV}\approx3\times10^{44}$~erg~s$^{-1}$) Compton-thick AGN; i.e.,\ this
source is a Compton-thick quasar.\footnote{Here we define a quasar as
an AGN with $L_{\rm 2-10 keV}\ge10^{44}$~erg~s$^{-1}$. This definition
is consistent with that derived from the classical optical quasar
threshold of $M_{\rm B}<-23$ (e.g.,\ Schmidt \& Green 1983), assuming
the $\alpha_{\rm OX}$ relationship of Steffen \etal (2006) and the
composite quasar spectrum of Vanden Berk \etal (2001); this definition
is also consistent with $L_{\rm X,*}$ derived for X-ray unobscured AGNs
(see Table~5 of Hasinger \etal 2005).} We then use these
X-ray--optical--mid-IR diagnostics to identify a further six distant
Compton-thick quasars in the literature and place limits on the space
density of Compton-thick quasars at $z\approx$~2. We adopted
$H_{0}=71$~km~s$^{-1}$~Mpc$^{-1}$, $\Omega_{\rm M}=0.27$, and
$\Omega_{\Lambda}=0.73$ throughout.

%
\section{Observations}
%

HDF-oMD49 (optical position $\alpha_{2000}=$~12$^{\rm h}$ 37$^{\rm m}$
04\fs 34, $\delta_{2000}=$~$+62^\circ$14$^{\prime}46\farcs2$; Steidel
et~al. 2002) was originally identified in the Lyman-Break Galaxy (LBG)
searches of Steidel \etal (2002; 2003). With a redshift of $z=2.211$,
HDF-oMD49 lies at the low-redshift end of the LBG population (median
redshift $z\approx$~3). HDF-oMD49 is also identified as a BX/BM galaxy,
based on the criteria of Steidel \etal (2004), and it is dubbed BM1156
in the BX/BM study of Reddy \etal (2006). The apparent contradiction in
HDF-oMD49 being selected as both an LBG and BX/BM is due to small
differences in the photometric data used in these studies.

%
%
\begin{figure*}
\centerline{\includegraphics[angle=0,width=9.0cm]{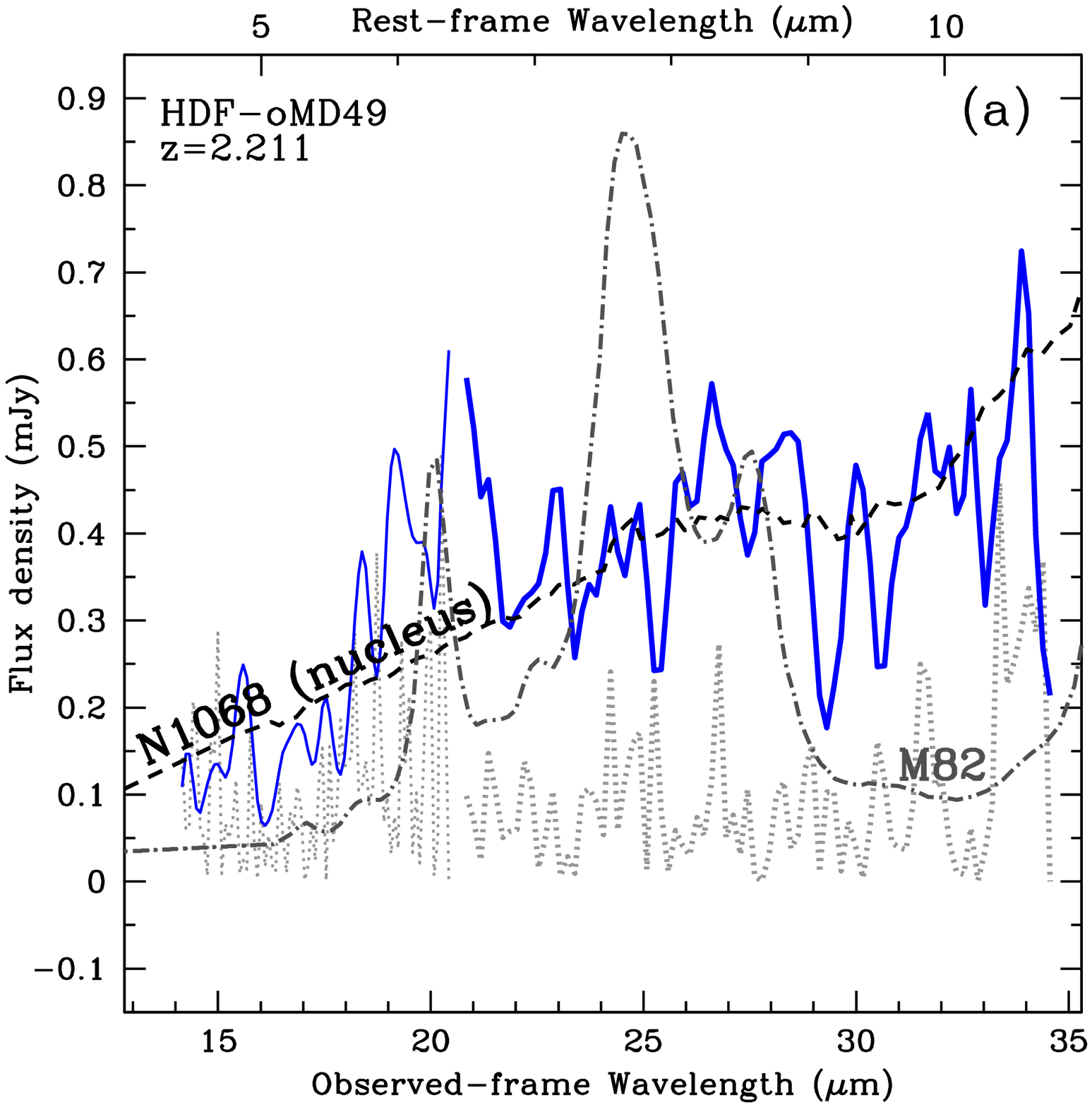}\hfill
\includegraphics[angle=0,width=9.0cm]{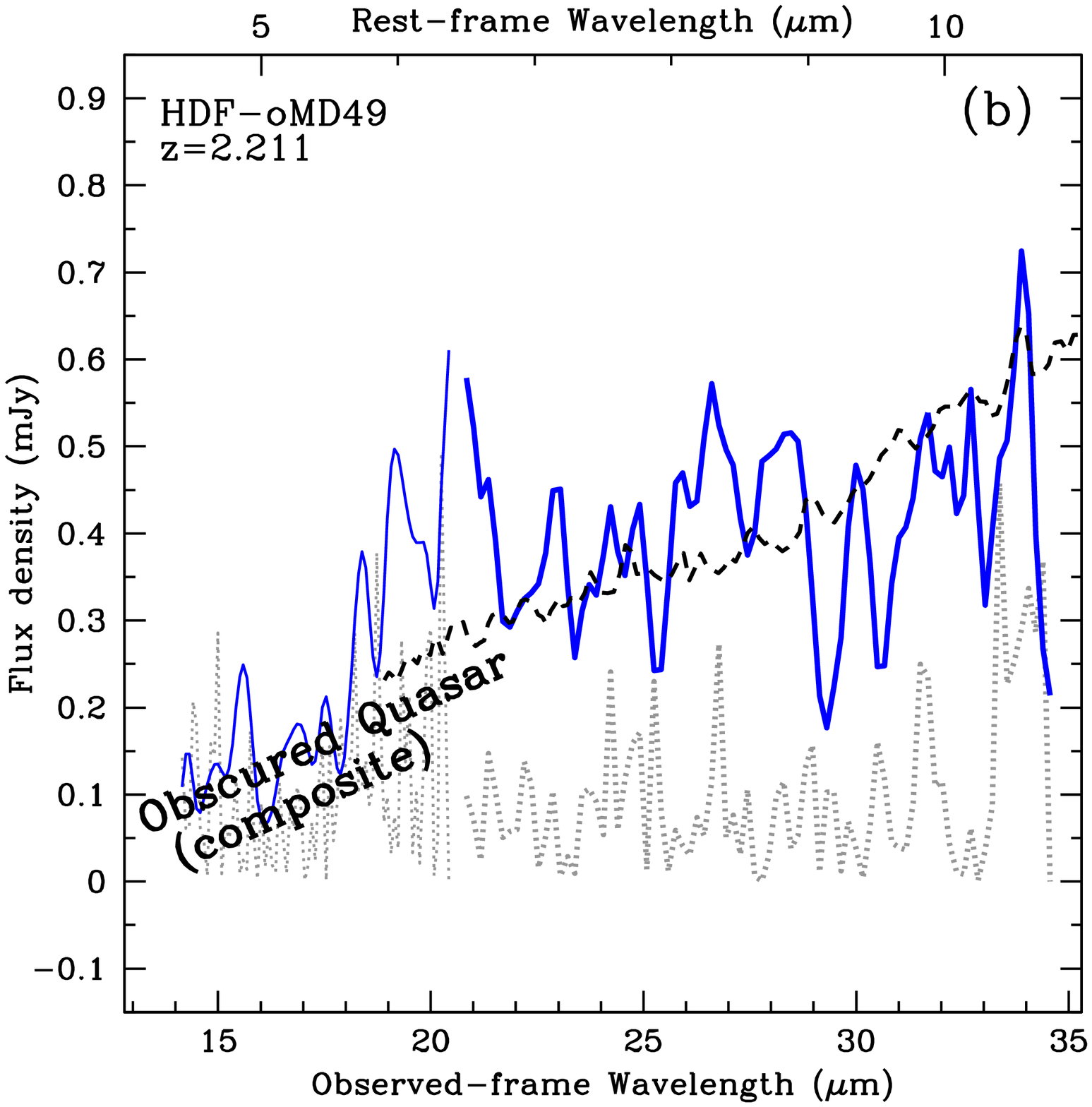}}
 \vspace{0.05in} \figcaption{{\it Spitzer}-IRS spectrum of HDF-oMD49
(solid line) for the LL1 (thick line) and LL2 (thin line) orders; the
1~$\sigma$ error arrays are plotted as a dotted line. Comparison
spectra (a: the starburst galaxy M82 and the Compton-thick AGN
NGC~1068; b: the composite spectrum of X-ray obscured Compton-thin
quasars from Sturm et~al. 2006) are shifted to $z=2.211$ and overlaid
on the data of HDF-oMD49. The mid-IR spectrum of HDF-oMD49 is
consistent with that of an AGN; see \S3.1 for quantitative
constraints.}
\end{figure*}

\vspace{0.1in}

\subsection{Spitzer Data}

HDF-oMD49 is detected by IRAC at 3.6--8.0~$\mu$m and MIPS at 24~$\mu$m
in the {\it Spitzer} observations obtained as part of the Great
Observatories Origins Deep Survey (GOODS) legacy project
(P.I.~M.~Dickinson; R.~Chary et~al., in prep.); the 3.6~$\mu$m,
4.5~$\mu$m, 5.8~$\mu$m, 8.0~$\mu$m, and 24~$\mu$m fluxes are
6.1~$\mu$Jy, 7.0~$\mu$Jy, 14.8~$\mu$Jy, 40.1~$\mu$Jy, and 380~$\mu$Jy,
respectively. Although HDF-oMD49 is bright at 24~$\mu$m, it is
undetected in the ultra-deep 70~$\mu$m observations of Frayer
et~al. (2006; 3~$\sigma$ upper limit of $f_{\rm 70\mu m}<2.0$~mJy),
indicating that it has a ``warm'' mid-IR Spectral Energy Distribution
(SED; e.g.,\ Papovich et~al. 2007). HDF-oMD49 is also detected at
16~$\mu$m with $f_{\rm 16\mu m}=222$~$\mu$Jy using the ``peak-up''
imaging capability of {\it Spitzer}-IRS (H.~Teplitz et~al., in prep.).

The {\it Spitzer}-IRS spectroscopy of HDF-oMD49 was obtained in
spectral staring mode as part of program GO2-20456 (P.I.~R.~Chary)
using the Long-Low modules (LL; \hbox{$R\approx$~64--128;} Houck
et~al. 2004). HDF-oMD49 was observed with LL1 (20.5--37.9~$\mu$m) for 3
hrs, and additional observations with LL1 (0.93 hrs) and LL2 (1.33 hrs;
14.3--21.2~$\mu$m) were obtained when HDF-oMD49 fell serendipitously
into the slit of another target from the same program. We analyzed the
2D basic calibrated data files from the S14.0.0 {\it Spitzer}-IRS
pipeline data; data reduction includes cleaning rogue pixels, fitting
and removing the latent charge build-up, subtracting the sky, and
averaging the 2D files together. Spectral extraction was performed
using a 2 pixel wide window in SPICE and the data were calibrated using
the same extraction window on a standard-star spectrum.\footnote{See
http://ssc.spitzer.caltech.edu/postbcd/spice.html for details of
SPICE.} The average RMS of the LL1 and LL2 spectra are 97~$\mu$Jy and
120~$\mu$Jy, respectively. The {\it Spitzer}-IRS fluxes measured
through the 16~$\mu$m and 24~$\mu$m filter spectral response curves are
$180\pm120$~$\mu$Jy and $370\pm100$~$\mu$Jy, respectively, consistent
with the observed fluxes in these bands.\footnote{The spectral response
curves for the individual {\it Spitzer} instruments can be obtained
from http://ssc.spitzer.caltech.edu/obs/} For more details on the
observations and data reduction see Pope et al.~(2008).

\subsection{Chandra Data}

HDF-oMD49 is undetected in the published 2~Ms {\it Chandra} catalogs of
Alexander et~al. (2003a). However, these catalogs were produced using a
conservative source-detection algorithm to minimize the number of
spurious detections, and it is possible to identify robustly fainter
sources when searching for X-ray emission associated with a known
source (e.g.,\ see \S3.4.2 of Alexander et~al. 2003a and \S5 of
Alexander et~al. 2001). Indeed, adopting a {\sc wavdetect} (Freeman
et~al. 2002) false-positive probability threshold of $10^{-5}$ we
detected significant X-ray emission (X-ray position
$\alpha_{2000}=$~12$^{\rm h}$ 37$^{\rm m}$ 04\fs 32,
$\delta_{2000}=$~$+62^\circ$14$^{\prime}46\farcs4$) within 0.25~arcsec
of the optical position of HDF-oMD49 in the 0.5--8~keV,
\hbox{0.5--2~keV,} and 4--8~keV bands with $8.1\pm3.5$, $6.7\pm2.8$,
and $4.9\pm2.6$ counts, respectively (see also Laird
et~al. 2006). HDF-oMD49 remains undetected in all of the other four
X-ray bands explored by Alexander et~al. (2003a) and we calculated
3~$\sigma$ upper limits following \S3.4.1 of Alexander
et~al. (2003a). Although the S/N of the X-ray data is low, the low
background of {\it Chandra} allows sources with very low count rates to
be reliably detected. For example, with a measured background in the
detection aperture of HDF-oMD49 ($\approx$~2 pixel radius) of only
$\approx$~1 counts in the 0.5--2~keV band, the probability of detecting
$\approx$~7 counts by chance is $\approx10^{-5}$; furthermore,
HDF-oMD49 is detected in two independent bands, increasing the overall
probability that the detection is real.

The X-ray counts correspond to fluxes of
$7.1\times10^{-17}$~erg~s$^{-1}$~cm$^{-2}$ (0.5--8~keV),
$2.6\times10^{-17}$~erg~s$^{-1}$~cm$^{-2}$ (0.5--2~keV), and
\hbox{$1.5\times10^{-16}$~erg~s$^{-1}$~cm$^{-2}$} (4--8~keV), for
$\Gamma=1.4$ (i.e.,\ the spectral slope of the XRB). The detection in
the 4--8~keV band (rest-frame 12.8--25.7~keV) is particularly notable
since {\it Chandra} is significantly less sensitive at \hbox{4--8~keV}
than at \hbox{$<4$~keV} energies, suggesting that HDF-oMD49 has a flat
X-ray spectral slope; the non detection in the wider \hbox{2--8~keV}
band provides support for this hypothesis. The X-ray band ratio (the
2--8~keV/0.5--2~keV count-rate ratio) is $<1.9$, which corresponds to
$\Gamma>0.2$ and is consistent with the X-ray spectral slope derived
from the 4--8~keV/0.5--2~keV count-rate ratio ($\Gamma\approx$~0.4);
the difference in the X-ray fluxes estimated using $\Gamma=0.4$ instead
of $\Gamma=1.4$ is negligible given the low S/N of the X-ray data
($\simlt$~15\%). The corresponding rest-frame 1.6--6.4~keV,
6.4--12.8~keV, and 12.8--25.7~keV luminosities (or 3~$\sigma$ upper
limit) of HDF-oMD49 are $10^{42}$~erg~s$^{-1}$,
$<4\times10^{42}$~erg~s$^{-1}$, and $6\times10^{42}$~erg~s$^{-1}$,
respectively.

%
%

\begin{deluxetable*}{lcc|ccccccc|cc|l}
\tablecaption{Robustly Identified $z\approx$~2 Compton-thick Quasars}
\tablehead{
\colhead{} &
\colhead{} &
\colhead{$d_L$}       &
\colhead{$L_{\rm X, obs}$}&
\colhead{$\nu$$L_{\rm 6\mu m}$}&
\colhead{$L_{\rm Ly{\alpha}}$} &
\colhead{$L_{\rm CIV}$}         &
\colhead{$L_{\rm HeII}$}        &
\colhead{$L_{\rm CIII]}$}       &
\colhead{$L_{\rm [OIII]}$}      &
\colhead{$L_{\rm X, 6\mu m}$}  &
\colhead{$L_{\rm X, lines}$} &
\colhead{} \\
\colhead{Name}                 &
\colhead{$z$}                  &
\colhead{(Mpc)}                &
\colhead{(erg~s$^{-1}$)}              &
\colhead{(erg~s$^{-1}$)}              &
\colhead{(erg~s$^{-1}$)}              &
\colhead{(erg~s$^{-1}$)}              &
\colhead{(erg~s$^{-1}$)}              &
\colhead{(erg~s$^{-1}$)}              &
\colhead{(erg~s$^{-1}$)}              &
\colhead{(erg~s$^{-1}$)}              &
\colhead{(erg~s$^{-1}$)}              &
\colhead{Refs}}
\startdata
SMM~J123600+621047$^{a}$ & 2.002 & 15750 & $<42.4$ & 45.3$^b$ & \dots & \dots &\dots &\dots & \dots & 44.7 & \dots & 1, 2\\
HDF-oMD49$^{a}$ & 2.211 & 17800 & 42.1 & 45.2$^b$ & 42.5 & 42.4 &41.9 & 41.7 & \dots & 44.6 & 44.4 & 3, 4\\
FSC~10214+4724$^{c}$ & 2.285 & 18550 & 42.3 & 44.9$^b$ & \dots & \dots & \dots&\dots & 42.6 & 44.3 & 44.4 & 5, 6, 7\\
SW~J104406+583954 & 2.430 & 19990 & $<43.4$ & 45.8$^b$ & 43.9 & 43.5 & 42.9 & \dots & \dots & 45.2 & 45.6 & 8, 9\\
BX160$^{a}$  & 2.462 & 20290 & $<42.3$ & 44.9$^d$ & 42.3 & 41.8 & \dots & \dots&\dots & 44.3 & 44.1 & 3, 4\\
BX1637$^{a}$ & 2.487 & 20570 & $<42.7$ & 44.8$^d$ & 42.3 & \dots & \dots &\dots &\dots & 44.3 & 44.1 & 3, 4\\
SW~J104409+585224 & 2.540 & 21100 & $<43.4$ & 46.4$^b$ & 43.8 & 43.1 &\dots& \dots & \dots & 45.9 & 45.5 & 8, 10
\enddata

\vspace{0.05in} 
\tablenotetext{}{Notes -- The luminosities are given in logarithmic
units of erg~s$^{-1}$. The X-ray luminosities are given in the
rest-frame 2--10~keV band: $L_{\rm X, obs}$ is the observed X-ray
luminosity, calculated from the observed-frame 0.5-2~keV luminosity assuming
$\Gamma=1.4$, $L_{\rm X, 6\mu m}$ is the X-ray luminosity implied from
the rest-frame 6~$\mu$m AGN continumm, and $L_{\rm X, lines}$ is the
average X-ray luminosity implied from the emission-line luminosities;
see \S3.3 and \S4.1 for more details. The references correspond to the
X-ray, mid-IR, and optical emission-line data -- 1: Alexander
et~al. (2005b); 2: Pope et~al. (2008); 3: This paper; 4: Reddy
et~al. (2006); 5: Alexander et~al. (2005a); 6: Teplitz et~al. (2006);
7: Serjeant et~al. (1998); 8: Polletta et~al. (2006); 9: Weedman
et~al. (2006); 10: Polletta et~al. (2008).}
\tablenotetext{a}{Object lies in the GOODS-N field.}
\tablenotetext{b}{Rest-frame 6~$\mu$m luminosity is calculated from the
best-fitting AGN component to the {\it Spitzer}-IRS data.}
\tablenotetext{c}{Data have been corrected for an assumed lensing boost
of 50.}
\tablenotetext{d}{Rest-frame 6~$\mu$m luminosity is calculated from the
{\it Spitzer} SED.}

\end{deluxetable*}

%
\section{Results}
%

\subsection{The {\it Spitzer} SED}

In Fig.~1a we show the {\it Spitzer}-IRS spectrum of HDF-oMD49. The
spectrum is noisy, particularly in the LL2 order. However, there is no
significant emission from the dominant rest-frame 7.7~$\mu$m Polycyclic
Aromatic Hydrocarbon (PAH) feature found in star-forming galaxies,
indicating that the mid-IR spectrum does not have a strong contribution
from star-formation activity. By contrast, the {\it Spitzer}-IRS
spectrum of HDF-oMD49 is similar to the nearby Compton-thick AGN
NGC~1068 and even shows weak evidence for Si absorption at 9.7~$\mu$m,
a feature often seen in obscured AGNs (e.g.,\ Shi et~al. 2006; Hao
et~al. 2007; Spoon et~al. 2007).  The {\it Spitzer}-IRS spectrum of
HDF-oMD49 is also qualitatively similar to the composite mid-IR
spectrum of the X-ray obscured quasars ($L_{\rm
X}>10^{44}$~erg~s$^{-1}$; $N_{\rm H}>10^{22}$~cm$^{-2}$) investigated
by Sturm et~al. (2006); see Fig.~1b.

%
%
\begin{figure}[!t]
\includegraphics[angle=0,width=85mm]{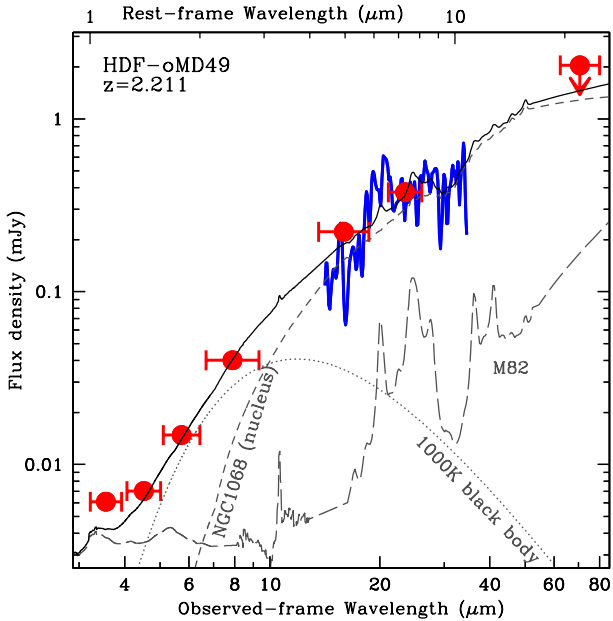}
\caption{Infrared data of HDF-oMD49, showing the {\it Spitzer}-IRS
spectroscopy (thick solid curve) and {\it Spitzer} photometry (solid
circle) at 3.6--70~$\mu$m. The error bar in the $x$-axis direction
indicates the width that corresponds to $>50$\% of the maximum
transmission in each of the photometric bands; see Footnote 3. The data
are compared to the SEDs of the Compton-thick AGN NGC~1068 (dashed
curve) and the starburst galaxy M82 (long-dashed curve), which are
shifted to $z=2.211$ and normalized based on the results from fitting
the mid-IR spectrum of HDF-oMD49 (see \S3.1). In order to fit the data
in the IRAC bands it was necessary to also add a black-body component
of very hot 1000~K dust (dotted curve). The thin solid curve shows the
combined model results. The data for the NGC~1068 and M82 SEDs were
taken from Rigopoulou et~al. (1999), Chary \& Elbaz (2001), Galliano
et~al. (2003), and Magnelli et~al. (2007).}
\end{figure}

We can quantify the contributions from AGN and star-formation activity
to the mid-IR luminosity of HDF-oMD49 by fitting the {\it Spitzer}-IRS
spectrum with AGN and star-formation templates. Adopting the approach
of Pope et~al. (2008), we fit the {\it Spitzer}-IRS spectrum with an
M82 starburst component and include an AGN component using the nuclear
spectrum of NGC~1068, performing a $\chi^2$ minimization to obtain the
best-fitting normalization. On the basis of this approach, we find that
the mid-IR spectrum of HDF-oMD49 is dominated by AGN activity, with a
limit of a 10\% contribution from star formation in the {\it
Spitzer}-IRS band; there is no clear signature for star formation and
this limit effectively corresponds to the noise per pixel in the
spectra. We get consistent results if we fit the mid-IR spectrum with
an M82 starburst component and an absorbed power-law component (to
represent the AGN), leaving the slope of the power law and the
extinction as free parameters. The best-fitting results with this
second approach give a spectral slope of $\alpha=$~2 (consistent with
that found for AGN-dominated sources; e.g.,\ Alonso-Herrero
et~al. 2003, where $f_{\nu}\approx\nu^{-{\alpha}}$) with
$\tau_{9.7{\mu}m}\approx$~0.65 and a $\approx$~5\% limit on the
contribution from star formation.

In Fig.~2 we show the 3.6--70~$\mu$m SED of HDF-oMD49 and plot on the
NGC~1068 and M82 templates, normalized to the values obtained from the
mid-IR spectral fitting. The strong observed-frame 3.6--8~$\mu$m
emission and the power-law-like featureless SED strengthens our
conclusion from the mid-IR spectral fitting, showing that the IR SED of
HDF-oMD49 is dominated by AGN activity. The strongly rising continuum
at $>4.5$~$\mu$m (rest-frame $>1.4$~$\mu$m) also indicates that the
contribution from starlight (which typically peaks at rest-frame
1.6~$\mu$m) is weak at these wavelengths. The combination of the
NGC~1068 and M82 templates gives a respectable fit to the overall SED
of HDF-oMD49 even though the wavelength coverage of the SED data is
three times broader than that used when fitting the mid-IR spectrum
(rest-frame 1.1--21.8~$\mu$m versus rest-frame
4.4--10.6~$\mu$m). However, the combination of the NGC~1068 and M82
components lies below the observed emission in the IRAC bands
(rest-frame $<3$~$\mu$m). This excess emission can be fitted by adding
a black-body component of very hot dust ($\approx$~1000~K), as often
found in obscured AGNs (e.g.,\ Alonso-Herrero et~al. 2001). On the
basis of the fitted NGC~1068 component, the rest-frame 6~$\mu$m
luminosity from HDF-oMD49 is $\nu$$L_{\rm 6\mu
m}\approx1.3\times10^{45}$~erg~s$^{-1}$ ($f_{\rm 6\mu m}=210$~$\mu$Jy),
indicating that the AGN is powerful (e.g.,\ compare to the nearby AGNs
explored by Lutz et~al. 2004).

Our analyses show that the IR SED of HDF-oMD49 is dominated by AGN
activity and allow for just a small contribution from star
formation. Using the best fitting star-formation component
normalizations and the relationship between PAH luminosity and
star-formation rate (SFR) given by Pope et~al. (2008; see Eqn.~7), we
predict SFR~$\approx$~60--120~$M_{\odot}$~yr$^{-1}$, with the range
representing the two different star-formation contributions we
estimated from the mid-IR spectral fitting. The radio detection of
HDF-oMD49 with $f_{\rm 1.4 GHz}\approx$~32~$\mu$Jy (G.~Morrison et~al.,
in prep.)  implies a SFR of $\approx$~500~$M_{\odot}$~yr$^{-1}$
(calculated following Bauer et~al. 2002). This is clearly inconsistent
with the mid-IR data, suggesting that either the radio emission is
dominated by the AGN or the radio-SFR relationship is not applicable
for high-redshift objects such as HDF-oMD49.

\subsection{The {\it Chandra} SED}

The contrast between the weak X-ray emission (see \S2.2) and the bright
AGN-dominated IR SED implies that the AGN in HDF-oMD49 is heavily
absorbed at X-ray energies. The signature of absorption should be
evident in the X-ray spectrum of HDF-oMD49. However, before examining
the \hbox{X-ray} data it is necessary to consider the contribution to
the X-ray emission from star formation. Assuming a SFR of
120~$M_{\odot}$~yr$^{-1}$ (see \S3.1) and the empirically determined
$L_{\rm X}$--SFR relationship found by Bauer et~al. (2002), we predict
a rest-frame 0.5--8~keV flux of
$2\times10^{-17}$~erg~s$^{-1}$~cm$^{-2}$, which corresponds to an
observed-frame 0.5--2~keV flux of
$1.1\times10^{-17}$~erg~s$^{-1}$~cm$^{-2}$ ($L_{\rm
X}\approx4\times10^{41}$~erg~s$^{-1}$), for a typical X-ray spectral
slope for star-forming galaxies of $\Gamma=1.8$. Consistent X-ray
luminosities are obtained from the empricially derived Ranalli \etal
(2003) relationship, although the Persic \etal (2004) relationship
would predict an $\approx$~4 times lower X-ray luminosity but only
takes into account the contribution from high-mass X-ray
binaries. These analyses show that the contribution from star formation
could be significant to the observed-frame 0.5--2~keV flux
($\approx$~40\%). However, the contribution will be negligible in the
other X-ray band in which HDF-oMD49 is detected (observed-frame
4--8~keV).

In Fig.~3 we show the X-ray data of HDF-oMD49 and compare them to the
AGN model SEDs adopted by Gilli et~al. (2007) to interpret the XRB. The
X-ray data of HDF-oMD49 clearly indicate that the AGN is absorbed at
X-ray energies; on the basis of the X-ray band ratio (see \S2.2), the
X-ray data are consistent with $N_{\rm H}>10^{23}$~cm$^{-2}$ (e.g.,\
see Fig.~4 of Alexander \etal 2003b). However, with our low S/N X-ray
data we cannot unambiguously distinguish between Compton-thick and
Compton-thin absorption. Furthermore, the X-ray spectrum of a
Compton-thick AGN is source specific and depends upon the strength of
the reflected and scattered X-ray components, in addition to the
underlying power-law continuum and any additional absorption; for
example, see Appendix~A of Guainazzi et~al. (2005) for information on
the difficulty of identifying Compton-thick AGNs on the basis of X-ray
colors alone. Therefore, without detailed X-ray spectral signatures
that clearly indicate the presence of Compton-thick absorption (e.g.,\
the identification of a prominent Fe~K emission line or a strong
reflection component), it is not possible to argue {\it solely} on the
basis of low S/N X-ray data that an AGN is absorbed by Compton-thick
material. In order to confirm whether such an X-ray weak source is
Compton thick, it is necessary to estimate the intrinsic AGN
luminosity.

%
%
\begin{figure}[!t]
\includegraphics[angle=0,width=85mm]{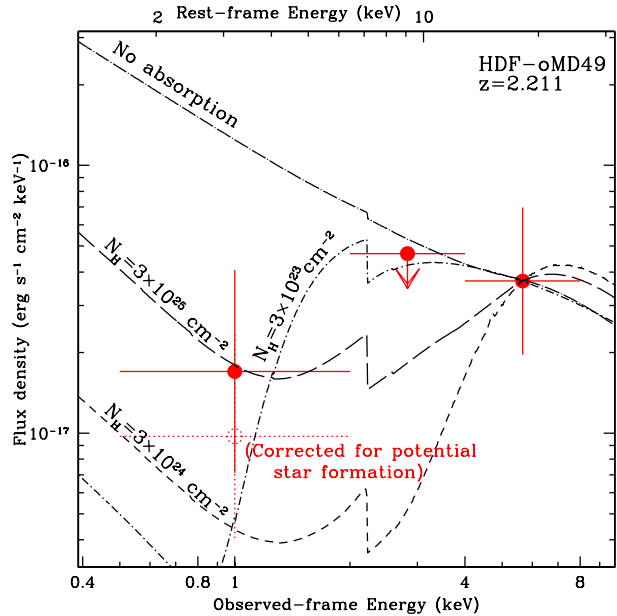}
\caption{X-ray data of HDF-oMD49, compared to the AGN model SEDs
adopted by Gilli et~al. (2007) to interpret the XRB (model $N_{\rm H}$
indicated) and normalized to the observed-frame 4--8~keV flux. The open
circle shows the X-ray emission at observed-frame 0.5--2~keV corrected
for the potential contribution from star formation; see \S3.2. The
X-ray data of HDF-oMD49 are consistent with those expected for a
heavily absorbed AGN, however, the low S/N X-ray spectrum alone
provides limited diagnostics to distinguish between Compton-thin and
Compton-thick absorption. We stress that since the data in this figure
is normalised to the observed 4--8~keV flux it does not demonstrate the
$>1$ order of magnitude difference in flux between similarly powerful
Compton-thin and Compton-thick AGNs; see Fig.~4.}
\end{figure}

%
%
\begin{figure*}[!t]
\centerline{\includegraphics[angle=0,width=125mm]{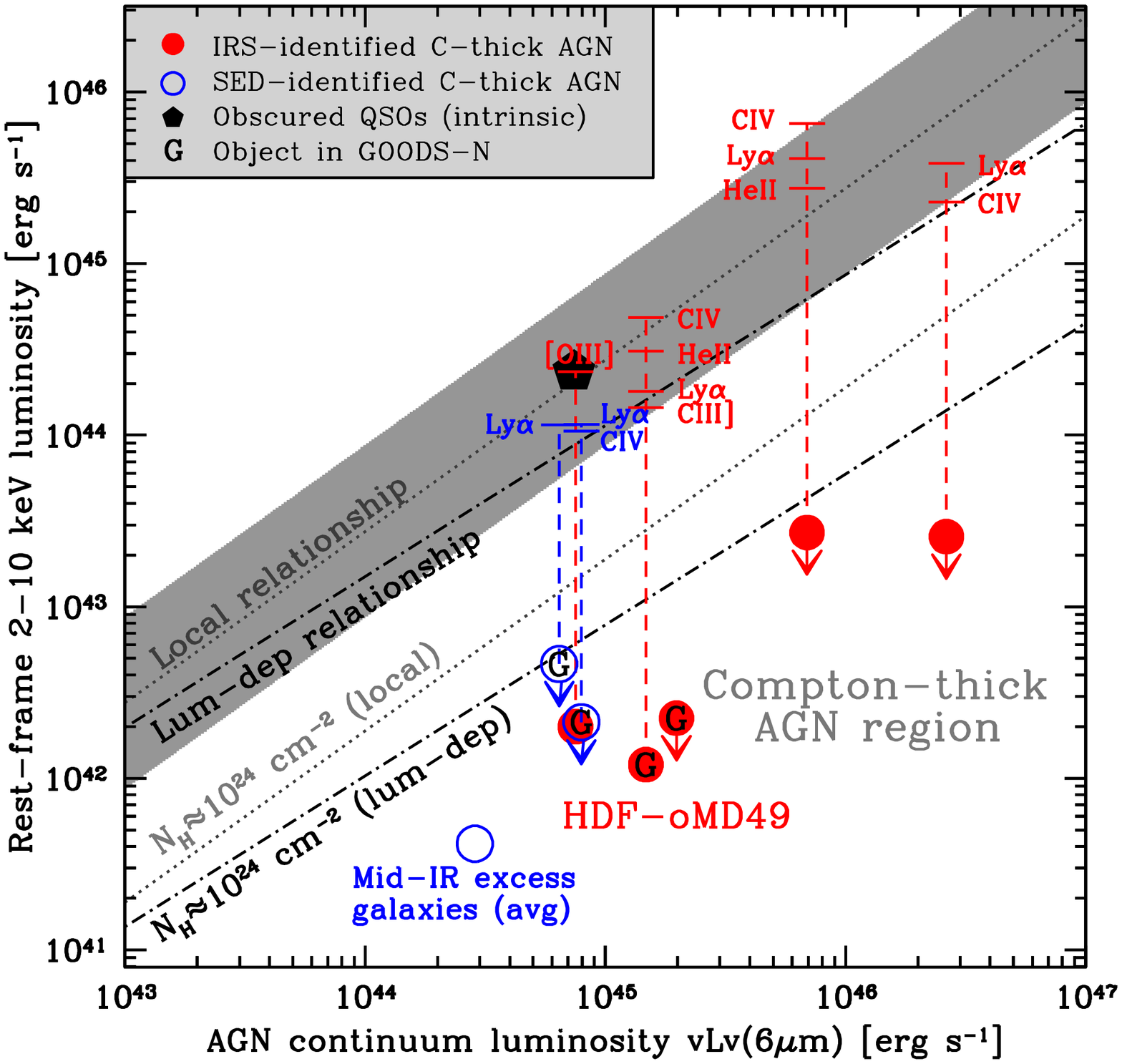}}
\caption{Rest-frame 2--10~keV luminosity versus 6~$\mu$m AGN continuum
luminosity. The observed and intrinsic properties of the $z\approx$~2
AGNs selected from our study are shown; the X-ray data are converted
from the observed-frame 0.5--2~keV band to the rest-frame 2--10~keV
band assuming $\Gamma=1.4$. Five of the objects are identified using
{\it Spitzer}-IRS and {\it Chandra} data (filled circle) and two are
identified using {\it Spitzer} photometry and {\it Chandra} data (open
circle); see Table~1. Objects indicated with a ``G'' lie in the GOODS-N
field. The 6~$\mu$m AGN luminosity is calculated from the best-fitting
AGN component for the objects with {\it Spitzer}-IRS spectroscopy, and
it is taken from the full SED for the objects with {\it Spitzer}
photometry. The filled pentagon shows the {\it intrinsic}
X-ray--6~$\mu$m luminosity ratio of the X-ray obscured quasars studied
by Sturm \etal (2006). The shaded region shows the range of intrinsic
X-ray--mid-IR luminosity ratios found for AGNs in the local Universe
(taken from Lutz \etal 2004), and the dotted lines indicate the
observed rest-frame 2--10~keV--6~$\mu$m luminosity ratio expected for a
typical AGN that is absorbed by $N_{\rm H}\approx10^{24}$~cm$^{-2}$
(using the model presented in Alexander et~al. 2005b); the
dashed-dotted region shows the relationship that would be predicted
assuming the luminosity-dependent $L_{\rm X}$--$L_{\rm mid-IR}$
relationship of Maiolino \etal (2007). All of the plotted AGNs lie
below the track for $N_{\rm H}\approx10^{24}$~cm$^{-2}$, suggesting
that they are Compton thick.  The dashed lines indicate the intrinsic
2--10~keV luminosities estimated from optical/near-IR spectroscopy (the
individual emission-line luminosities are labelled) and provide good
agreement with the intrinsic luminosities predicted from the mid-IR
spectroscopy, confirming that all of the plotted AGNs are Compton
thick. The average observed X-ray--IR luminosity ratio found for the
mid-IR excess galaxies explored by Daddi et~al. (2007a; open circle) is
also shown, indicating that their average properties are consistent
with those expected for Compton-thick AGNs about an order of magnitude
less luminous than those explored here (assuming 100\% of their objects
are Compton-thick AGNs; see \S4.4 for discussion); however,
spectroscopic observations of individual objects (some of which are
presented here) are required to provide robust confirmation of
Compton-thick AGN activity.}
\end{figure*}

\subsection{Quantifying the Intrinsic Properties of the AGN in HDF-oMD49}

A good proxy for the intrinsic luminosity of AGN activity is the
rest-frame mid-IR emission, which can provide an absorption-independent
measure of the intrinsic luminosity even in the presence of
Compton-thick absorption (e.g.,\ Krabbe \etal 2001; Lutz \etal 2004;
Maiolino \etal 2007; Horst \etal 2008). When combined with the observed
X-ray emission, the rest-frame mid-IR luminosity can therefore provide
a diagnostic for the presence of heavy absorption. In this study we
adopt the X-ray--6~$\mu$m luminosity relationship found by Lutz
et~al. (2004), which was determined by constraining the AGN continuum
component in large-aperture mid-IR spectra for a sample of nearby
AGNs. We favour the Lutz \etal (2004) X-ray--mid-IR luminosity
relationship over the small-aperture mid-IR imaging studies of Krabbe
\etal (2001) and Horst \etal (2008) since HDF-oMD49 is unresolved by
{\it Spitzer}. Furthermore, the Lutz \etal (2004) study covers a
broader range in luminosity, and includes AGNs as luminous as those
explored here. Indeed, we note that the X-ray--mid-IR luminosity ratio
found for X-ray obscured but Compton-thin quasars (Sturm et~al. 2006)
is consistent with the Lutz et~al. (2004) relationship, indicating that
it is applicable for the objects explored in our study; see
Fig.~4. However, the differences in the X-ray--mid-IR luminosity of
these studies are small [the Lutz \etal (2004) luminosity ratio is
within a factor of $\approx$~2--3 of that found by Krabbe \etal (2001)
and Horst \etal (2008), assuming typical SEDs (i.e.,\ NGC~1068;
Mrk~231; M82) when converting to a common rest-frame wavelength of
6~$\mu$m].

Since it is now well established that the X-ray-to-optical luminosity
ratio of optically selected AGNs is luminosity dependent (e.g.,\
Vignali \etal 2003; Steffen \etal 2006), it is also interesting to
explore whether the X-ray--mid-IR luminosity ratio is luminosity
dependent. Krabbe \etal (2001), Lutz \etal (2004), and Horst \etal
(2008) did not find evidence for a luminosity dependent X-ray--mid-IR
luminosity ratio for local AGNs over $\approx$~3--5 orders of magnitude
in luminosity. By comparison, the inference from the study of distant
optically selected quasars by Maiolino \etal (2007) is that the
X-ray--mid-IR luminosity ratio {\it may} be luminosity dependent. For
example, solving Eqn.~1 and Eqn.~5 in Maiolino \etal (2007) gives an
X-ray--mid-IR luminosity relationship of $L_{\rm X}$~$\propto$~$L_{\rm
6.7{\mu} m}$$^{0.88}$. However, there is considerable uncertainty in
this result, and the derived X-ray--mid-IR luminosity relationship
would be linear if a different X-ray--optical luminosity ratio from
Steffen \etal (2006) is used (i.e.,\ if Eqn.~1b is adopted instead of
Eqn.~1c). In any case, the differences between the Lutz \etal (2004)
and Maiolino \etal (2007) relationships are small (factor $<$~2--3) for
the luminosities of the majority of the AGNs explored here ($L_{\rm
X}<3\times10^{44}$~erg~s$^{-1}$; see Fig.~4); however, we indicate the
cases where our results would significantly change if the luminosity
dependent version of the Maiolino \etal (2007) X-ray--mid-IR luminosity
relationship is used.

In Fig.~4 we show the rest-frame 2--10~keV luminosity versus 6~$\mu$m
continuum luminosity of HDF-oMD49 and compare it to those found for AGN
in the local Universe; we converted the rest-frame 6.7~$\mu$m data used
by Maiolino \etal (2007) to rest-frame 6~$\mu$m assuming the SED of
NGC~1068, which results in small corrections ($\approx$~10\%). The
rest-frame 2--10~keV luminosity of HDF-oMD49 ($L_{\rm
X}=1.4\times10^{42}$~erg~s$^{-1}$) is estimated from the observed-frame
0.5--2~keV luminosity (rest-frame 1.6--6.4~keV), for a spectral slope
of $\Gamma=1.4$. Assuming both the X-ray--mid-IR luminosity
relationship found for local AGNs (Lutz et~al. 2004), and the
relationship derived from Maiolino et~al. (2007), the intrinsic
rest-frame \hbox{2--10~keV} luminosity of HDF-oMD49 is
$>10^{44}$~erg~s$^{-1}$, suggesting that the X-ray absorption toward
the AGN in HDF-oMD49 is Compton thick; see Fig.~4 and Table~1.
However, to reduce further uncertainties on the X-ray--IR luminosity
relationship due to, for example, the geometry and covering factor of
the gas and dust surrounding the X-ray emitting region, it is useful to
have other measurements of the intrinsic AGN luminosity.

The rest-frame UV spectrum of HDF-oMD49 shows strong AGN emission lines
(CIV, CIII], HeII, Ly$\alpha$; see Fig.~2 of Steidel et~al. 2002),
which can be used to provide additional estimates of the intrinsic AGN
luminosity. We calculated the emission-line fluxes using the Steidel
\etal (2002) optical spectrum and integrating the flux over the
emission-line profiles; see Reddy et~al. (2006) for more
details. Adopting the average emission-line ratios of Netzer
et~al. (2006), to calculate the expected [OIII]$\lambda$5007
luminosity, and the [OIII]$\lambda$5007--X-ray flux ratio found by
Mulchaey et~al. (1994), the range of predicted intrinsic rest-frame
\hbox{2--10~keV} luminosity from the different emission lines is
\hbox{(1.8--4.8)}~$\times10^{44}$~erg~s$^{-1}$. We get similar
intrinsic X-ray luminosities if we use the [OIII]$\lambda$5007--X-ray
flux ratio found by Alonso-Herrero \etal (1997) and we get higher
intrinsic X-ray luminosities (up to an order of magnitude higher for
the most luminous AGNs) if we use the Netzer \etal (2006)
[OIII]$\lambda$5007--X-ray flux ratio. We have not corrected these
emission-line fluxes for extinction or contamination from star
formation, which are poorly constrained at rest-frame UV
wavelengths. However, we note that the Mulchaey et~al. (1994),
Alonso-Herrero \etal (1997), and Netzer \etal (2006) correlations were
also derived using uncorrected emission-line fluxes. The uncertainties
in both the emission-line ratios and the [OIII]$\lambda$5007--X-ray
flux ratio are $\approx$~2 (see Table~1 of Netzer \etal 2006 and
Table~2 of Mulchaey \etal 1994), giving a combined uncertainty of a
factor of $\approx$~3, comparable to the uncertainty in the intrinsic
X-ray luminosity based on the rest-frame 6~$\mu$m luminosity. The
intrinsic rest-fame 2--10~keV luminosities predicted from both the
optical and the mid-IR spectroscopy are in good agreement, providing
compelling evidence for Compton-thick absorption of an intrinsically
luminous quasar in HDF-oMD49 ($L_{\rm
2-10~keV}\approx3\times10^{44}$~erg~s$^{-1}$; $N_{\rm
H}\gg10^{24}$~cm$^{-2}$).

An alternative scenario for the weak X-ray emission from HDF-oMD49 is
that the AGN has recently faded (e.g.,\ Guainazzi \etal 1998).
However, since the light-crossing time of the mid-IR emitting region is
likely to be short (i.e.,\ on the basis of the dust-sublimation radius
it will be $\approx$~1--2~yrs; e.g.,\ see Table~1 of Granato, Danese,
\& Franceschini 1997 and Fig.~30 of Suganuma \etal 2006), this scenario
is only tenable for HDF-oMD49 if we have caught the AGN during a very
brief transitionary stage where the X-ray emission has faded but the
mid-IR emission is still strong. The good agreement between the fluxes
measured from the {\it Spitzer}-MIPS photometry and the {\it
Spitzer}-IRS spectroscopy, taken 2~years apart, suggests that this
scenario is unlikely.

%
\section{Discussion}
%

We have used {\it Spitzer}-IRS spectroscopy, 3.6--70~$\mu$m photometry,
optical spectroscopy, and ultra-deep 2~Ms {\it Chandra} data to show
that HDF-oMD49 hosts an intrinsically luminous AGN that is heavily
obscured by Compton-thick material ($L_{\rm
2-10~keV}\approx3\times10^{44}$~erg~s$^{-1}$; $N_{\rm
H}\gg10^{24}$~cm$^{-2}$); i.e.,\ it is a Compton-thick quasar. The S/N
of the X-ray data alone was too poor to show conclusively that the AGN
is obscured by Compton-thick material. However, when combined with
rest-frame 6~$\mu$m and optical emission-line luminosity constraints,
the evidence for Compton-thick absorption in this X-ray faint source is
compelling. Additional support for this interpretation comes from the
similarity of the mid-IR spectrum of HDF-oMD49 and the composite mid-IR
spectrum of X-ray obscured but Compton-thin quasars; see Fig.~1b.

On the basis of the comparatively weak ultra-hard \hbox{4--8}~keV
emission from HDF-oMD49, even the rest-frame \hbox{12.8--25.7}~keV
emission appears to be heavily suppressed. For example, assuming an
X-ray spectral slope of $\Gamma=2.0$, the predicted
\hbox{12.8--25.7}~keV luminosity based on the intrinsic rest-frame
2--10~keV luminosity is a factor of $\approx$~20 higher than the
observed 12.8--25.7~keV luminosity
($\approx1.3\times10^{44}$~erg~s$^{-1}$ versus
$\approx6\times10^{42}$~erg~s$^{-1}$; see \S2.2). It is difficult to
relate this factor of $\approx$~20 suppression to an accurate estimate
of the absorbing column density since the observed X-ray emission in
the 12.8--25.7~keV band could be dominated by reflected and scattered
components (e.g.,\ NGC~1068; Matt \etal 1997; see Fig.~8.1 in Comastri
2004). However, on the basis of the {\sc plcabs} model (Yaqoob 1997) in
{\sc xspec}, which considers the effects of electron scattering and
photoelectric absorption through an absorbing medium, this amount of
suppression in the 12.8--25.7~keV band would correspond to an absorbing
column density of $N_{\rm H}\approx4\times10^{24}$~cm$^{-2}$. Due to
the absence of a reflection component in this model, and the
simplifying assumption of a spherical geometry, this absorbing column
density should only be considered representative (e.g.,\ see Fig.~8.1
in Comastri 2004 for an alternative model).

In this final section we adopt the approach that we used in \S3 to
identify other distant Compton-thick AGNs using data obtained from the
literature. Utilizing these constraints we then directly estimate the
space density of Compton-thick quasars at $z\approx$~2 and compare our
results to current observational and theoretical constraints.

\subsection{The Identification of other Distant Compton-thick AGNs}

From a search of the published {\it Spitzer}-IRS spectroscopic data,
there are four good candidate $z\approx$~2 Compton-thick AGNs that have
mid-IR spectroscopy and the {\it essential} sensitive X-ray constraints
required to identify Compton-thick AGN activity: FSC~10214+4714,
SW~J104406+583954, SW~J104409+585224, and SMM~J123600+621047.

FSC~10214+4714 is a strongly lensed Seyfert 2 galaxy at $z=2.285$ which
produces luminous 6~$\mu$m AGN continuum emission (Teplitz et~al. 2006)
but is only weakly detected at X-ray energies, despite an effective
lensing-corrected {\it Chandra} exposure of $\approx$~2~Ms (Alexander
et~al. 2005a). In Fig.~4 we plot FSC~10214 and provide secondary
support for the intrinsic luminosity of the AGN using the
[OIII]$\lambda$5007 luminosity from Alexander et~al. (2005a) and the
[OIII]$\lambda$5007--X-ray flux ratio of Mulchaey et~al. (1994); see
Table~1. The combination of the X-ray data and optical--mid-IR
spectroscopy confirms that FSC~10214 hosts a Compton-thick quasar
($L_{\rm 2-10 keV}\approx2\times10^{44}$~erg~s$^{-1}$), even given the
considerable scatter in the relationships.

SW~J104406+583954 and SW~J104409+585224 are X-ray weak candidate
Compton-thick AGNs identified in the {\it Chandra}/SWIRE survey
(Polletta \etal 2006) that lie at at $z=2.430$ and $z=2.540$,
respectively. The optical and mid-IR spectra of both sources show clear
signatures of AGN activity (Polletta \etal 2006, 2008; Weedman \etal
2006), which we use to determine the intrinsic luminosities of the
central sources; see Table~1. In Fig.~4 we plot their X-ray--6~$\mu$m
luminosities, showing that the properties of both objects are
consistent with those expected for a luminous Compton-thick quasar
($L_{\rm 2-10 keV}>10^{45}$~erg~s$^{-1}$); however, we note that the
evidence is weaker for SW~104406 if the luminosity dependent
X-ray-to-mid-IR luminosity relationship inferred from Maiolino
et~al. (2007) is used.

SMM~J123600+621047 is a $z=2.002$ submillimeter emitting galaxy (SMG)
with a mid-IR-bright AGN (Pope et~al. 2008) that is undetected in the
2~Ms CDF-N data (Alexander et~al. 2003a); we ran {\sc wavdetect} using
a false-positive probability threshold of $10^{-5}$ but did not detect
X-ray emission from this source (see Table~1 of Alexander \etal
2005b). The X-ray--6~$\mu$m luminosity ratio indicates that it hosts a
Compton-thick AGN (see Fig.~4), although optical and near-IR
spectroscopic observations do not reveal the signatures of AGN activity
(e.g.,\ Swinbank et~al. 2004; Chapman et~al. 2005), leaving some
uncertainty on the intrinsic luminosity of the AGN. However, since the
mid-IR luminosity of the AGN in SMMJ~123600 is about an order of
magnitude larger than that found in typical SMGs (e.g.,\ Lutz
et~al. 2005; Men{\'e}ndez-Delmestre et~al. 2007; Valiante et~al. 2007;
Pope et~al. 2008), the $\approx$~10 times larger intrinsic X-ray
luminosity derived here seems plausible ($L_{\rm X, 6\mu
m}\approx5\times10^{44}$~erg~s$^{-1}$; i.e.,\ compare to Alexander
et~al. 2005b); see Table~1. In Alexander \etal (2005c), it was argued
that SMGs represent a rapid black-hole growth phase before the onset of
optically luminous quasar activity. The discovery of this one mid-IR
luminous X-ray undetected SMG in the CDF-N field increases current
constraints on the {\it integrated} black-hole growth density in SMGs
by a factor of $\approx$~2 (see Fig.~2 of Alexander et~al. 2005c).

Other studies have used optical spectroscopy to identify AGNs that are
mid-IR bright and X-ray weak/undetected but lack {\it Spitzer}-IRS
spectroscopy. Reddy et~al. (2006) found two $z\approx$~2.4--2.5
optically identified AGNs (BX160; BX1637) that are X-ray undetected in
the 2~Ms CDF-N survey and have power-law like IR SEDs [the 24~$\mu$m
fluxes of these sources ($f_{\rm 24\mu m}\approx$~100--140~$\mu$Jy) are
too faint for good S/N {\it Spitzer}-IRS spectroscopy]; see
Table~1. Following our approach for HDF-oMD49 (see \S2.2), we ran {\sc
wavdetect} using a false-positive probability threshold of $10^{-5}$
but did not detect X-ray emission from either source; 3~$\sigma$ X-ray
upper limits are calculated following \S3.4.1 in Alexander
et~al. (2003a). We plot the X-ray-to-6~$\mu$m luminosity ratios of
these objects in Fig.~4. The mid-IR data, when combined with the
optical emission-line luminosities, indicates that these sources are
likely to be Compton-thick quasars; we determined their emission-line
properties using the same procedure as for HDF-oMD49 (see \S3.3). The
evidence would be weaker for BX1637 if the luminosity dependent
X-ray--mid-IR luminosity ratio inferrred from Maiolino et~al. (2007) is
used, although we note that the X-ray--6~$\mu$m luminosity ratio for
this source is an upper limit.

We also note that other studies have selected candidate Compton-thick
AGNs using a combination of multi-wavelength data with sensitive X-ray
constraints (e.g.,\ Donley et~al. 2005, 2007; Alonso-Herrero 2006;
Daddi et~al. 2007a; Mart{\'{\i}}nez-Sansigre et~al. 2007; Fiore \etal
2008); many other studies have identified obscured AGNs using mid-IR
spectroscopy but lack the {\it essential} X-ray data to provide a case
for Compton-thick absorption. However, none of these studies has
optical or mid-IR spectroscopy to provide accurate measurements of the
intrinsic AGN luminosities.

%
%
\begin{figure}[!t]
\centerline{\includegraphics[angle=0,width=85mm]{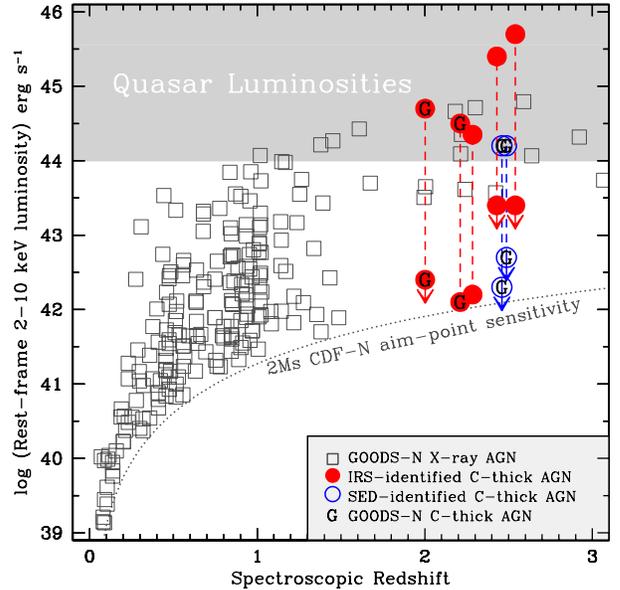}}
\caption{Rest-frame 2--10~keV luminosity versus spectroscopic redshift
for the X-ray detected sources in the GOODS-N region of the CDF-N
survey and the seven Compton-thick AGNs identified here; the four
Compton-thick AGNs that lie in the GOODS-N field are indicated with a
``G''. The dotted curve shows the aim-point sensitivity limit for the
CDF-N survey; see Fig.~18 of Alexander et~al. (2003a) for the drop in
{\it Chandra} sensitivity for off-axis sources. The source redshifts
are predominantly taken from Barger et~al. (2003). The dashed lines
connect the observed X-ray luminosity of the Compton-thick AGNs to
their intrinsic luminosities; the intrinsic luminosities correspond to
the mean luminosity estimated from the combination of the 6~$\mu$m
luminosity and the average emission-line luminosity given in
Table~1. Although the four Compton-thick AGNs that lie in the GOODS-N
region are either X-ray weak or undetected, they are amongst the most
intrinsically luminous AGNs in the GOODS-N field.}
\end{figure}

\vspace{0.2in}
\subsection{The Properties of Distant Compton-thick AGNs}

%
%
\begin{figure*}
\centerline{\includegraphics[angle=0,width=9.0cm]{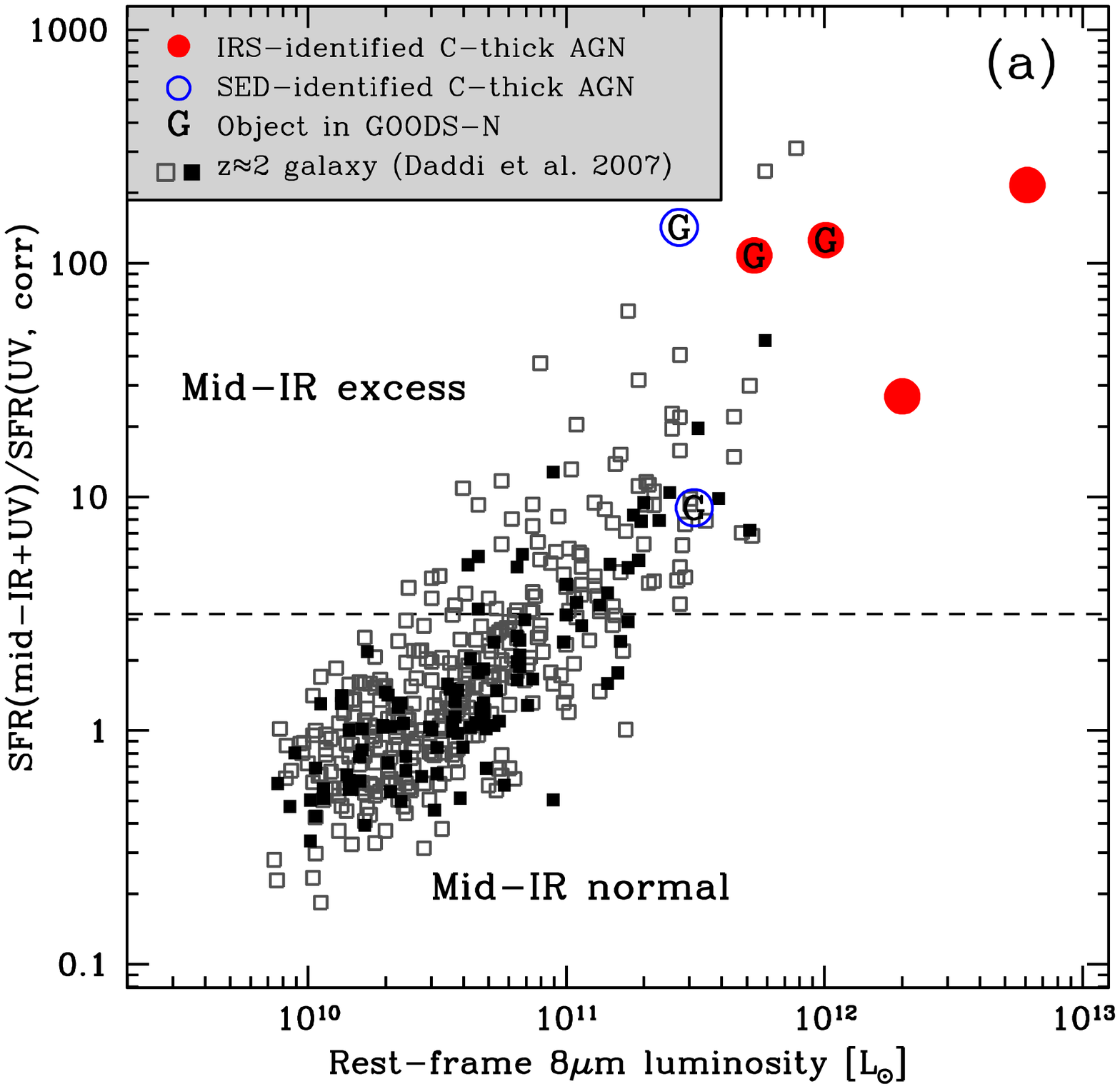}\hfill
\includegraphics[angle=0,width=9.0cm]{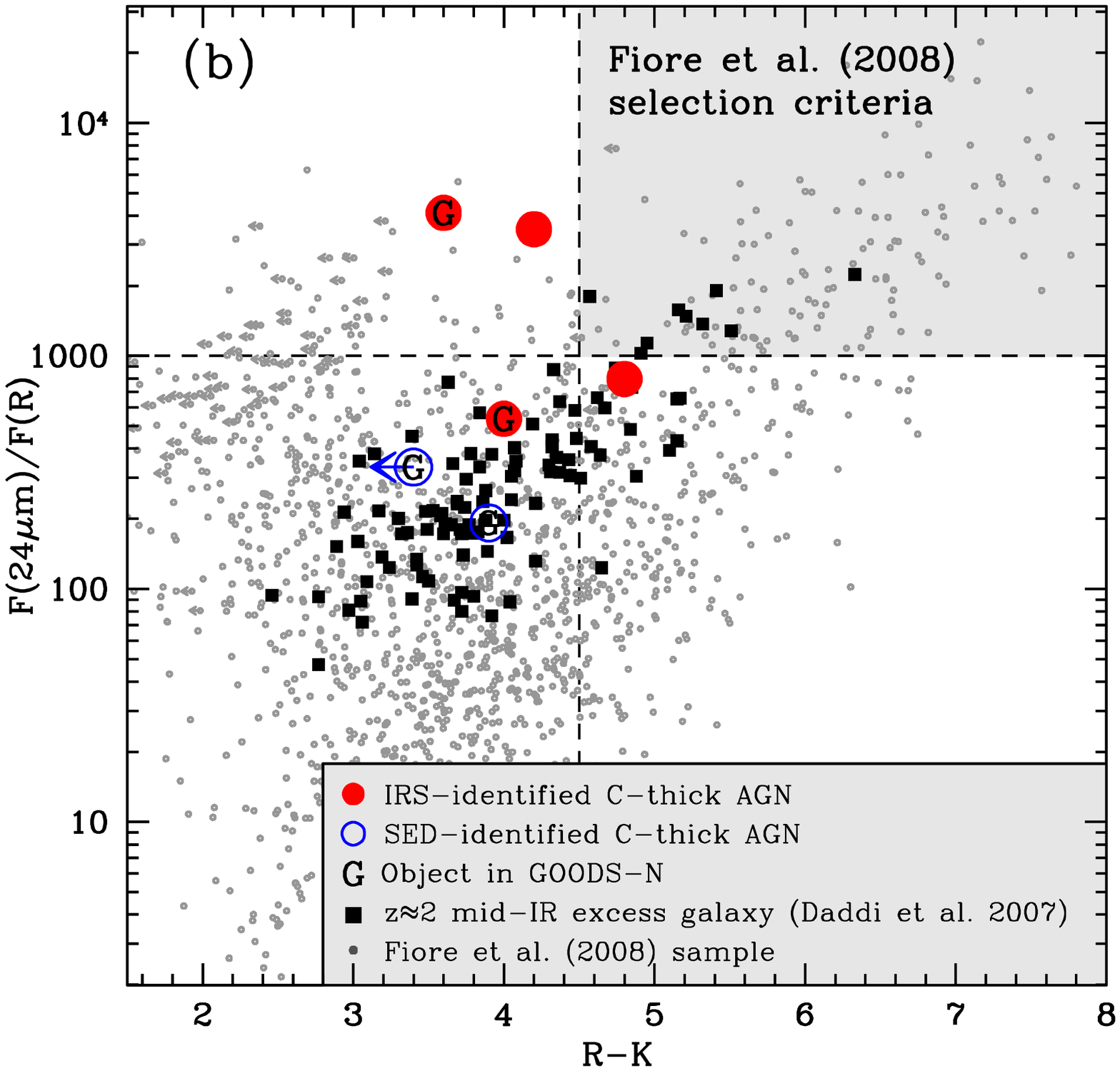}}
\caption{Selection of distant Compton-thick AGNs. (a) ratio of
mid-IR-based to UV-based star-formation rate (SFR) versus 8~$\mu$m
luminosity for the $z\approx$~2 galaxies studied by Daddi \etal
(2007a,b) and the spectroscopically identified Compton-thick quasars;
the open and filled symbols correspond to objects with photometric and
spectroscopic redshifts, respectively. This figure has been adapted
from Fig~2 of Daddi \etal (2007a). (b) mid-IR--optical flux ratio
versus $R-K$ color for the galaxies investigated by Fiore \etal (2008)
and Daddi \etal (2007a,b) and the spectroscopically identified
Compton-thick quasars; see also Fig.~3 of Fiore \etal
(2008). FSC~10214+4724 is not plotted on these figures since there is
some uncertainty on the lensing magnification of the star-forming
component. The dashed lines indicate the selection criteria for the
Daddi \etal (2007a) and Fiore \etal (2008) studies. Our
spectroscopically identified Compton-thick quasars are extreme examples
of the mid-IR excess galaxy population of Daddi \etal (2007a) but none
of them formally match the Fiore \etal (2008) selection criteria, which
is probably due to the requirement for an optical signature of AGN
activity in our sample. This implies that there could be many more
distant Compton-thick AGNs than that suggested by either study alone.}
\end{figure*}

The intrinsic X-ray--mid-IR continuum ratios, inferred using the
rest-frame UV emission-line luminosities of the six Compton-thick
quasars with optical AGN signatures, are consistent with those found for
AGNs in the local Universe; see Fig.~4. This is in agreement with the
results of Sturm \etal (2006), who found that the {\it average}
intrinsic X-ray--6~$\mu$m luminosity of distant X-ray obscured quasars
are similar to those found for local AGNs, suggesting that the objects
explored here are the extreme Compton-thick end of the \hbox{X-ray}
obscured quasar population. We also showed in \S3.1 and Fig.~1b that
the {\it Spitzer-IRS} spectrum of HDF-oMD49 is similar to the composite
mid-IR spectrum produced by Sturm \etal (2006), providing a
particularly strong case for a quasar that is absorbed in the X-ray
band by Compton-thick material in this source. These pieces of evidence
also suggest that the accretion properties of distant optically
identifiable AGNs are similar to those observed in local AGNs, in
agreement with previous studies that have explored the X-ray--optical
and X-ray--infrared properties of optically selected AGNs (e.g.,\
Steffen et~al. 2006; Jiang et~al. 2006).

In Fig.~5 we show the observed and intrinsic X-ray luminosities versus
redshift of the seven Compton-thick quasars identified here. Four of
the Compton-thick quasars lie in the GOODS-N region
($\approx$~160~arcmin$^2$) of the CDF-N and, despite being X-ray
weak/undetected, they are amongst the most intrinsically luminous AGNs
identified in this field. This shows that even in the deepest X-ray
survey currently available, we are only able to directly probe the peak
of the Compton-thick AGN population at $z\approx$~2. The two luminous
Compton-thick quasars were identified in the wider-area SWIRE survey
($\approx$~0.6~deg$^2$) and are amongst the most luminous obscured AGNs
known.

Four of the seven Compton-thick quasars are detected in the X-ray band,
most typically at $>2$~keV. Three of these objects have hard X-ray
spectra ($\Gamma<1$; HDF-oMD49; SW~104406; SW~104409; Polletta \etal
2006) and one object has a comparatively soft X-ray spectrum
($\Gamma\approx$~1.6; FSC~10214; Alexander \etal 2005c). The
comparatively soft X-ray spectrum of FSC~10214 might be dominated by
scattered emission and star formation from the host galaxy, as often
found in Compton-thick AGNs (e.g.,\ Bassani et~al. 1999; Comastri 2004;
Guainazzi et~al. 2005). We can also place basic X-ray spectral
constraints on the three X-ray undetected Compton-thick quasars by
stacking their X-ray data; all of these objects lie in the GOODS-N
field where the X-ray data are particularly sensitive. Using the X-ray
stacking code adopted by Worsley \etal (2005), we obtain
$\approx$~3~$\sigma$ detections in the 0.5--8~keV and 2--8~keV bands,
and have a $\approx$~2~$\sigma$ constraint in the 0.5--2~keV band; the
significance of these stacking results were calculated using 10,000
Monte Carlo trials. The stacked data corresponds to fluxes (and a
3~$\sigma$ upper limit) of $2.8\times10^{-16}$~erg~s$^{-1}$~cm$^{-2}$
(0.5--8~keV), $<2.3\times10^{-17}$~erg~s$^{-1}$~cm$^{-2}$ (0.5--2~keV),
and $2.7\times10^{-16}$~erg~s$^{-1}$~cm$^{-2}$ (2--8~keV). The average
rest-frame 2--10~keV luminosity constraint for these three objects
derived from the observed-frame 0.5--2~keV flux is
$<10^{42}$~erg~s$^{-1}$. The derived X-ray spectral slope of
$\Gamma<0.3$ unambiguously confirms that heavily obscured AGNs are
present in these sources (e.g.,\ see Fig.~2a in Alexander \etal 2005b).

\subsection{The Selection of Distant Compton-thick AGNs}

The seven Compton-thick quasars investigated here were selected on the
basis of unambiguous evidence for AGN activity from optical and/or
mid-IR spectropscopy. The advantage with this approach is that we have
been able to estimate the intrinsic AGN properties of these objects
using the optical--mid-IR spectroscopy. However, the requirement for
spectroscopic evidence of AGN activity has led to a comparatively small
sample of objects. Nevertheless, we can provide some insight into the
completeness of Compton-thick AGN selection by comparing the properties
of our reliably identified Compton-thick quasars to the larger samples
of candidate Compton-thick AGNs that have been photometrically
identified in other studies. Arguably, the most complete identification
studies of distant Compton-thick AGNs to date are Daddi \etal (2007a,b)
and Fiore \etal (2008).

Daddi \etal (2007a,b) identified a large population of X-ray undetected
$z\approx$~2 galaxies with an excess of mid-IR emission over that
expected from star formation (as predicted using the dust-extinction
corrected UV luminosity). X-ray stacking analyses revealed a hard X-ray
spectral slope from these objects ($\Gamma=0.8$), comparable to that of
the Compton-thick quasars and unambiguously identifying the presence of
heavily obscured AGNs; see \S4.2. In Fig.~6a we plot the ratio of
mid-IR based to UV-based star-formation rate (SFR) versus 8~$\mu$m
luminosity of the Compton-thick quasars and compare them to Daddi \etal
(2007a,b); we have not calculated the properties of FSC~10214 due to
the significant uncertainties in the lensing magnification of the
star-forming regions. The Compton-thick quasars lie significantly above
the mid-IR excess threshold defined by Daddi \etal (2007a), indicating
that they are extreme examples of the mid-IR excess galaxy
population. This analysis shows reliably that at least a subsample of
the mid-IR excess galaxy population host Compton-thick AGNs.

Fiore \etal (2008) took a complementary approach to that of Daddi \etal
(2007a,b) and selected X-ray undetected objects with extreme
mid-IR-to-optical flux ratios ($f_{\rm 24\mu m}/f_{\rm R}>1000$) and
red optical colors ($R-K>4.5$). X-ray stacking analyses of these
objects revealed a hard X-ray spectrum with a slope similar to that
found for the mid-IR excess galaxies ($\Gamma\approx$~1.0, calculated
from their stacked X-ray count rates). In Fig.~6b we plot the
mid-IR-to-optical flux ratio versus $R-K$ color of the Compton-thick
quasars and compare them to the candidate Compton-thick AGNs studied by
Fiore \etal (2008) and Daddi \etal (2007a); as before, we did not
calculate the properties of FSC~10214. Formally, none of the
Compton-thick quasars match the candidate Compton-thick AGN selection
criteria of Fiore \etal (2008), although two of the objects lie
close. Furthermore, only $\approx$~10\% of the mid-IR excess galaxies
of Daddi \etal (2007a) match the Fiore \etal (2008) criteria. These
analyses show that the candidate Compton-thick AGN criteria of Daddi
\etal (2007a) and Fiore \etal (2008) are mutually exclusive, suggesting
that the distant X-ray undetected AGN population may be larger than
either study has suggested. The requirement in our study for optical
spectroscopic redshifts and the optical identification of AGN activity
(with the exception of SMM~J123600) probably causes a bias against
selecting objects with red $R-K$ colors, which by definition will be
optically faint.

\subsection{The Ubiquity of Distant Compton-thick AGNs}

Our seven Compton-thick quasars do not comprise a complete
sample. However, four of the objects lie in the \hbox{GOODS-N} field,
providing basic constraints on the space density of Compton-thick
quasars at $z\approx$~2--2.5. On the basis of the comoving volume at
$z=$~2--2.5 in a region the size of the GOODS-N field, we estimate a
comoving space density for Compton-thick quasars with $L_{\rm 2-10
keV}\approx10^{44}$--$10^{45}$~erg~s$^{-1}$ of
$\Phi\approx$~(0.7--2.5)~$\times10^{-5}$~Mpc$^{-3}$, where the range
only reflects the uncertainty due to small-number statistics (e.g.,\
Gehrels 1986).

In Fig.~7 we plot the Compton-thick quasar space density and compare it
to that found in other studies. Our results suggest that the space
density of Compton-thick quasars is $\approx$~1--5 times higher than
that of comparably luminous unobscured quasars at $z\approx$~2--2.5
($\Phi\approx5\times10^{-6}$~Mpc$^{-3}$; e.g.,\ Hasinger \etal
2005). Since the Gilli \etal (2007) model of the X-ray background
constrains the number of Compton-thick quasars to be the same as the
number of unobscured quasars, our Compton-thick quasar space density is
also \hbox{$\approx$~1--5} times higher than the Compton-thick quasar
predictions of Gilli \etal (2007). As we mention in \S4.1 and show in
Fig.~4, the evidence for a Compton-thick quasar in BX1637 is weaker if
the X-ray--mid-IR luminosity ratio inferred by Maiolino \etal (2007) is
used. However, we note that a significantly more important issue is
likely to be the effect of cosmic variance since we have only
identified Compton-thick quasars in a small region over a narrow
redshift range. For example, within the GOODS-N field, there are also
four X-ray unobscured quasars that lie at $z\approx$~2--2.5, indicating
that the true Compton-thick--unobscured quasar ratio may be toward the
lower end of the value given above. Nevertheless, our results clearly
indicate that a large fraction of the growth of black holes was
obscured by Compton-thick material, in general agreement with results
based on less-reliable photometrically identified objects (e.g.,\ Daddi
\etal 2007a; Mart{\'{\i}}nez-Sansigre et al. 2007; Fiore \etal 2008),
the predictions made by theoretical models (e.g.,\ Marconi \etal 2004;
Treister \etal 2006), and the Compton-thick AGN fraction inferred by
the study of Maiolino et~al. (2007; Fig.~7a).

%
%
\begin{figure}[!t]
\centerline{\includegraphics[angle=0,width=9.0cm]{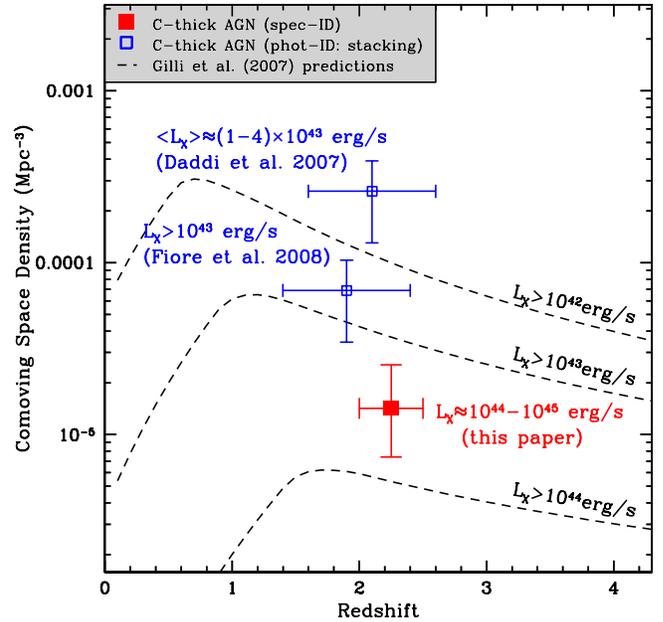}}
\caption{Space density of Compton-thick AGNs. The filled square
corresponds to the results determined here for spectroscopically
identified Compton-thick quasars with $L_{\rm 2-10
keV}\approx10^{44}$--$10^{45}$~erg~s$^{-1}$ that lie in the GOODS-N
region of the CDF-N field. The open squares correspond to the results
obtained by Daddi \etal (2007a) and Fiore \etal (2008) using
photometrically identified objects and X-ray stacking analyses; the
plotted values assume that 100\% of their objects are Compton-thick
AGNs and they are therefore upper limits (see \S4.4 for discussion). The
dashed lines are the predicted space densities of Compton-thick AGNs
for a range of X-ray luminosity lower limits from the Gilli \etal
(2007) XRB model. The space density of the Compton-thick quasars
robustly identified here is a factor $\approx$~1--5 higher than that
found for comparably luminous unobscured quasars, indicating that
Compton-thick accretion is ubiquitious at $z\approx$~2; see \S4.4 for a
more detailed discussion.}
\end{figure}

We can compare our estimated space density to those determined from the
photometric selection of candidate Compton-thick AGNs by Daddi \etal
(2007a) and Fiore \etal (2008). On the basis of the X-ray data and the
X-ray--6~$\mu$m luminosity, Daddi \etal (2007a) argued that their
candidate Compton-thick AGNs have an average intrinsic luminosity of
$L_{\rm 2-10 keV}\approx$~(1--4)~$\times10^{43}$~erg~s$^{-1}$. The
intrinsic luminosities of these mid-IR excess galaxies are about an
order of magnitude lower than the Compton-thick quasars identified
here, and consequently their space density is higher
($\Phi\approx2.6\times10^{-4}$~Mpc$^{-3}$; see Fig.~7). Fiore \etal
(2008) estimated the intrinsic X-ray luminosities of their X-ray
undetected candidate Compton-thick AGNs using the 5.8~$\mu$m
luminosities, finding $L_{\rm 2-10 keV}>10^{43}$~erg~s$^{-1}$. Taking
into account the estimated redshift range of their objects
($z\approx$~1.2--2.6), we calculate a space density of
$\Phi\approx7\times10^{-5}$~Mpc$^{-3}$; as mentioned in Fiore \etal
(2008), this space density is consistent with that predicted for
Compton-thick AGNs using the XRB model from Gilli \etal (2007). Taken
at face value, these results suggest that lower-luminosity
Compton-thick AGNs were \hbox{$\approx$~5--20} times more common than
Compton-thick quasars at $z\approx$~2. However, since the primary
evidence for Compton-thick AGN activity in these studies was from X-ray
stacking analyses, there could be significant numbers of
non-Compton-thick AGNs in their samples (i.e.,\ either Compton-thin
AGNs or star-forming galaxies). Both Daddi \etal (2007a) and Fiore
\etal (2008) asssessed the ``contamination'' from star-forming galaxies
to their stacked X-ray results and concluded that $\simgt$~50\% and
$\approx80\pm15$\% of the objects are heavily obscured AGNs,
respectively. However, these percentages refer to the fraction of
photometrically identified objects that are heavily obscured AGNs and
direct optical-mid-IR spectroscopic identification will be required to
determine what fraction of these AGNs are obscured by Compton-thick
material.

We can also compare our results to those determined for more luminous
Compton-thick quasars. Polletta \etal (2006) identified five candidate
Compton-thick quasars at \hbox{$z\approx$~1.4--2.6} with $L_{\rm 2-10
keV}>10^{45}$~erg~s$^{-1}$ on the basis of the {\it Chandra} data in
the SWIRE survey, giving a space density of
$\Phi\approx5.7\times10^{-7}$~Mpc$^{-3}$. Two of these objects have
been presented here as robust Compton-thick quasars (SW~J104406 and
SW~J104409; see Table~1). However, the other three candidate
Compton-thick quasars are all undetected at 24~$\mu$m ($f_{24\mu
m}<230$~$\mu$Jy; see Table 2 of Polletta \etal 2006). This suggests
that either the intrinsic AGN luminosities of these Compton-thick
quasars are more modest ($L_{\rm 2-10 keV}<10^{45}$~erg~s$^{-1}$) than
those estimated by Polletta \etal (2006) or that the X-ray--mid-IR
relationship shown in Fig.~4 does not hold for these luminous objects.
Mart{\'{\i}}nez-Sansigre et al.(2007) identified ten Compton-thick
quasars at $z\approx$~1.7--4.2 with an average bolometric luminosity of
$L_{\rm BOL}\approx10^{47}$~erg~s$^{-1}$, giving a space density of
$\Phi\approx3.9\times10^{-8}$~Mpc$^{-3}$. On the basis of the
bolometric conversion of Elvis \etal (1994), which
Mart{\'{\i}}nez-Sansigre et al.(2007) also adopted, the estimated
average X-ray luminosities are $L_{\rm
2-10~keV}\approx3\times10^{45}$~erg~s$^{-1}$ for these sources,
suggesting that they are very luminous AGNs. However, since two of
their objects are Compton-thin quasars that are detected at X-ray
energies, we can directly estimate the intrinsic 2--10~keV luminosities
of the candidate Compton-thick quasars using the average
X-ray--6~$\mu$m luminosity ratio of the Compton-thin quasars. Taking
this approach and using the median 24~$\mu$m flux for the sample, the
typical intrinsic X-ray luminosities of these quasars would be more
modest ($L_{\rm 2-10~keV}\approx3\times10^{44}$~erg~s$^{-1}$), casting
significant doubt on whether they are obscured by Compton-thick
material; the rest-frame 6~$\mu$m fluxes are derived from the observed
24~$\mu$m fluxes using the NGC~1068 SED to apply small
$K$-corrections. Direct spectroscopic identification of these
photometrically classified Compton-thick quasars is required to place
more direct constraints, which should be possible with current
instrumentation.

Lastly, we compare our results to the Compton-thick AGNs identified by
Tozzi \etal (2006) using X-ray spectral analysis of X-ray detected AGNs
in the {\it Chandra} Deep Field-South survey. Fourteen of the objects
investigated by Tozzi \etal (2006) had X-ray spectra that were better
fitted by a pure reflection model than an absorbed power-law model, and
were consequently classified as Compton-thick AGNs; derived space
densities are $\Phi\approx3\times10^{-5}$~Mpc$^{-3}$ at $z\approx$~1
and $\Phi\approx8\times10^{-6}$~Mpc$^{-3}$ at $z\approx$~2. While the
Tozzi \etal (2006) study provides the most quantitative X-ray
identification of distant Compton-thick AGNs to date, it is also
possible for Compton-thin AGNs to have a pure reflection component
(i.e.,\ if the power-law component has temporarily decreased), and
other pieces of evidence are required to confirm that these are
Compton-thick AGNs. Two of the seven objects with spectroscopic
redshifts have strong Fe~K$\alpha$ emission lines, confirming that they
are likely to be Compton thick, but supporting evidence for
Compton-thick absorption is lacking in the other objects.

\subsection{Prospects for Improved Observational Constraints}

The current study provides diagnostics to identify distant
Compton-thick AGNs but is limited in source statistics, covers a narrow
redshift range, and is restricted to the brightest AGNs at
$z\approx$~2. Future studies could focus on obtaining optical--mid-IR
spectroscopy of lower-luminosity $z\approx$~2 candidate Compton-thick
AGNs, as well as identifying objects at lower and higher
redshifts. Assuming the NGC~1068 template used here and the average
X-ray--mid-IR relationship shown in Fig.~4, an AGN with $L_{\rm 2-10
keV}\approx10^{43}$~erg~s$^{-1}$ will have 24~$\mu$m fluxes of
$\approx$~2~$\mu$Jy, $\approx$~15~$\mu$Jy, $\approx$~0.3~mJy, and
$\approx$~5~mJy at $z\approx4$, $z\approx2$, $z\approx0.7$, and
$z\approx0.2$, respectively; these fluxes will typically scale linearly
with luminosity (see \S3.3). Mid-IR bright objects ($f_{\rm 24\mu
m}>0.2$~mJy) should be identifiable with {\it Spitzer}-IRS but mid-IR
spectroscopic identification of fainter objects will need to wait until
the launch of the {\it James Webb Space Telescope}. Optical and near-IR
spectroscopy may be easier to obtain for many of the candidate
Compton-thick AGNs, although only if the AGNs have identifiable
emission lines, which could be weak in low-luminosity and dust-reddened
AGNs (e.g.,\ Caccianiga \etal 2007; see \S4.3).

Ultimately, deep X-ray data are required to search directly for the
presence of Compton-thick absorption in individual objects using high
S/N X-ray spectroscopy. For the majority of the objects investigated
here, this will require $\gg2$~Ms exposures with current X-ray
telescopes; see Fig.~5. However, due to its large lensing boost it is
possible to provide reasonable X-ray spectral constraints for
FSC~10214+4724 in a modest {\it Chandra} exposure (e.g.,\ a 100~ks {\it
Chandra} exposure would yield almost 100 X-ray counts, $>10$ times more
than found for HDF-oMD49; see Fig.~3). Large improvements in the X-ray
spectral constraints for distant Compton-thick AGNs will need to wait
for the next generation of X-ray observatories (e.g.,\ {\it
Constellation-X}, {\it NuSTAR}, {\it Simbol-X}, {\it XEUS}), where the
foci are high-resolution spectroscopy and/or high-energy imaging
($>10$~KeV) of faint X-ray sources. For example, a 1~Ms {\it XEUS}
exposure will yield $\approx$~5000 counts for HDF-oMD49, allowing
detailed X-ray spectral analyses and providing the potential to search
for changes in the X-ray emission due to column density variations,
giving direct insight into the environment around the black hole of a
distant Compton-thick AGN (e.g.,\ Risaliti \etal 2002).

%
\section{Conclusions}
%

We have presented {\it Spitzer}-IRS spectroscopy and
\hbox{3.6--70~$\mu$m} photometry of HDF-oMD49, a $z=2.211$ optically
identified AGN that is formally undetected in the 2~Ms \hbox{CDF-N}
observation. From a combination of optical--mid-IR spectroscopy and
X-ray data, we have shown that HDF-oMD49 hosts an intrinsically
luminous quasar that is obscured by Compton-thick material ($L_{\rm
2-10~keV}\approx3\times10^{44}$~erg~s$^{-1}$; $N_{\rm
H}\gg10^{24}$~cm$^{-2}$). We selected six further $z\approx$~2 AGNs
from the literature (four with {\it Spitzer}-IRS spectroscopy) and used
the same X-ray--optical--mid-IR diagnostics applied to HDF-oMD49 to
show that these objects are also likely to be Compton-thick quasars
with $L_{\rm 2-10 keV}>10^{44}$~erg~s$^{-1}$. We demonstrated that
these Compton-thick quasars would be classified as mid-IR excess
galaxies, on the basis of the Daddi \etal (2007a) definition, providing
the first spectroscopic confirmation of Compton-thick AGN activity in a
subsample of these $z\approx$~2 mid-IR bright galaxies. Four of these
Compton-thick quasars lie in the GOODS-N field, and we used these
objects to constrain the space density of distant Compton-thick
quasars, finding $\Phi\approx$~(0.7--2.5)~$\times10^{-5}$~Mpc$^{-3}$ at
$z\approx$~2--2.5.  On the basis of our results, Compton-thick quasars
were as ubiquitious as unobscured quasars at $z\approx$~2--2.5,
implying that a large fraction of the growth of supermassive black
holes must have been obscured by Compton-thick material.

\acknowledgements We acknowledge support from the Royal Society (DMA),
the {\it Spitzer Space Telescope} Fellowship program (AP), the Natural
Sciences and Engineering Research Council of Canada and the Canadian
Space Agency (AP), the {\it Chandra} Fellowship program (FEB), and the
NASA LTSA grant NAG5-13035 (WNB). We thank R.~Gilli for generously
providing his model tracks and Compton-thick AGN spectra, and for his
insightful comments on an earlier draft of this paper. We also thank
F.~Fiore for providing the data from his sample, E.~Sturm for providing
the composite obscured quasar spectrum, C.~Done, M.~Polletta and
E.~Treister for scientific feedback, and the referee for useful
suggestions. This work is based in part on observations made with the
{\it Spitzer Space Telescope}, which is operated by the Jet Propulsion
Laboratory, California Institute of Technology under a contract with
NASA. The IRS was a collaborative venture between Cornell University
and Ball Aerospace Corporation funded by NASA through the Jet
Propulsion Laboratory and Ames Research Center.



\end{document}